\documentclass{aa}
\usepackage{graphicx}
\usepackage{amssymb}
\usepackage{epsfig}
\usepackage{amsmath}

\newcommand{\dis}{\displaystyle}

\newcommand{\gap}{\gtrsim}
\newcommand{\lea}{\lesssim}
\newcommand{\reps}{\epsilon}
\newcommand{\rx}{x}

\newcommand{\tf}{t_{\rm f}}
\newcommand{\dd}{\delta}
\newcommand{\Ii}{\langle I^{\rm i}\rangle}
\newcommand{\Io}{\langle I^{\rm o}\rangle}
\newcommand{\brut}{\sqrt{1-\mu^2}}
\newcommand{\bruts}{{1-\mu^2}}
\newcommand{\asin}{{\rm asin}}
\newcommand{\acos}{{\rm acos}}
\newcommand{\vecr}{{\bf r}}
\newcommand{\vecz}{{\bf z}}

\begin{document}

   \title{X-ray line emission from a fragmented stellar wind}

   \author{A. Feldmeier \and L. Oskinova \and W.-R. Hamann}

   \offprints{A. Feldmeier}

   \institute{Astrophysik, Institut f\"ur Physik, Universit\"at
              Potsdam, Am Neuen Palais 10, 14469 Potsdam, Germany\\
              \email{afeld, lida, wrh@astro.physik.uni-potsdam.de}}

   \date{Received ; accepted }

   \abstract{We discuss X-ray line formation in dense O~star winds. A
       random distribution of wind shocks is assumed to emit X-rays
       that are partially absorbed by cooler wind gas. The cool gas
       resides in highly compressed fragments oriented perpendicular
       to the radial flow direction. For fully opaque fragments, we
       find that the blueshifted part of X-ray line profiles remains
       flat-topped even after severe wind attenuation, whereas the red
       part shows a steep decline. These box-type, blueshifted
       profiles resemble recent Chandra observations of the O3 star
       $\zeta$~Pup. For partially transparent fragments, the emission
       lines become similar to those from a homogeneous wind.
       \keywords{Stars: winds, outflows -- X-rays: stars -- Radiative
       transfer}}

   \maketitle

\section{Introduction}

Recent Chandra and XMM-Newton observations of O~stars show resolved
X-ray emission line profiles from H- and He-like ions. The line
profiles observed exhibit a variety of shapes: (1) the stars
$\theta^1$~Ori~C (Schulz et al.~2000) and $\zeta$~Ori (Waldron \&
Cassinelli 2001) show strongly wind-broadened, symmetric, and {\it
nonshifted} profiles; (2) $\delta$~Ori~A (Miller et al.~2002) and
$\tau$~Sco (Mewe et al.~2003) display weakly or nonbroadened,
symmetric, and nonshifted profiles; (3) $\zeta$~Pup (Cassinelli et
al.~2001; Kahn et al.~2001) shows strongly broadened, asymmetric, and
blueshifted profiles.

Strong wind-broadening should be a consequence of the lines forming in
the fast wind. Furthermore, symmetric and nonshifted profiles indicate
line formation in an extended and optically thin wind, whereas
asymmetric and blueshifted profiles indicate line formation in a dense
wind, where X-rays from the stellar back hemisphere (with respect to
the observer) are more strongly absorbed by cool intervening wind gas
than X-rays from the front hemisphere. The most puzzling result of the
above observations is that a putatively dense wind like $\zeta$~Ori's
shows line profiles suggestive of an optically thin wind.

In order to gain a better understanding of the above observations, we
address X-ray emission line formation in a structured stellar wind
with thin, aligned absorbing layers. We consider a wind consisting of
very dense, discrete fragments of gas shells that float through
essentially empty space and are oriented perpendicular to the radial
flow direction. Such a hydrodynamic structure seems plausible for
radial wind flows reaching Mach numbers of 100 and being subject to
strong perturbations. Radiative postshock cooling zones should lead
here to highly compressed, thin gas sheets lying perpendicular to the
flow direction.

More specifically, a wind with oriented absorber layers could result
from the combined action of the line deshadowing instability,
photospheric turbulence, and the Rayleigh-Taylor instability. The line
deshadowing instability compresses the initially homogeneous gas flow
from the photosphere into thin, highly overdense shells (Owocki et
al.~1988). In linear approximation, the line deshadowing instability
has no lateral component (Rybicki et al.~1990), and the resulting flow
structure obeys spherical symmetry, i.e.~is shell-like. However,
turbulence-induced velocity fluctuations at the wind base lead to
significant differences in the spacing of flow structure (shocks and
dense gas shells) along neighboring wind rays. Therefore, dense gas
patches form instead of extended shells. The Rayleigh-Taylor
instability should further fragment these patches.

We neglect lateral motions (e.g.~eddies), and assume purely radial
flow. The flow shall obey spherical symmetry in the statistical sense,
i.e., the spatial distribution of fragments depends on radius, but not
on latitude or azimuth. In the first part of the present paper, the
fragments are assumed to be fully opaque. This assumption is dropped
in the second part.

X-ray emission originates from shocks associated with the dense
fragments. Hydrodynamic simulations show that the shocks occur almost
exclusively on the inner, starward facing side of the fragments,
i.e.~are reverse shocks (Owocki et al.~1988). To remain general,
however, we leave it open whether strong forward shocks occur on the
outer fragment faces, too (Lucy 1982).

\section{Wind model}

\subsection{Assumptions}

Along the lines of this general description, we set up an idealized
stellar wind with highly compressed gas fragments oriented
perpendicular to the radial flow direction, making the following
assumptions:

\medskip

\noindent (1) The gas flow is purely radial, and the radial wind speed
$v(r)=v_\infty$ is constant.

\noindent (2) No gas resides outside an outer atmospheric radius
$r=1$. Inside a radius $\reps\ll 1$, the optical depth is infinite
due to the presence of the central star and dense gas above the
photosphere.

\noindent (3) Between radii $\reps$ and 1, all absorbing wind gas
resides in 2-D absorbers of high density, termed {\it fragments}, that
are oriented perpendicular to the radial flow direction.

\noindent (4) The opening angle of an individual fragment as seen from
the star is infinitesimal and independent of $r$.

\noindent (5) The fragments are uniformly random distributed as
function of the spherical coordinates $r$, $\theta$, and $\phi$.

\noindent (6) On average, $N$ fragments occur along each radial ray.

\noindent (7) Each fragment is fully opaque. (This assumption will be
dropped in Sect.\,\ref{nonopaque}.)

\noindent (8) X-rays are emitted from within a sphere of radius
$\rx\le 1$, with the emissivity scaling as $\eta\sim r^{-2m}$, where
typically $m=2$ (emission $\sim$ density squared). There is no X-ray
reemission by the absorbing fragments.

\medskip

\begin{figure}
\centering
\includegraphics[width=6cm]{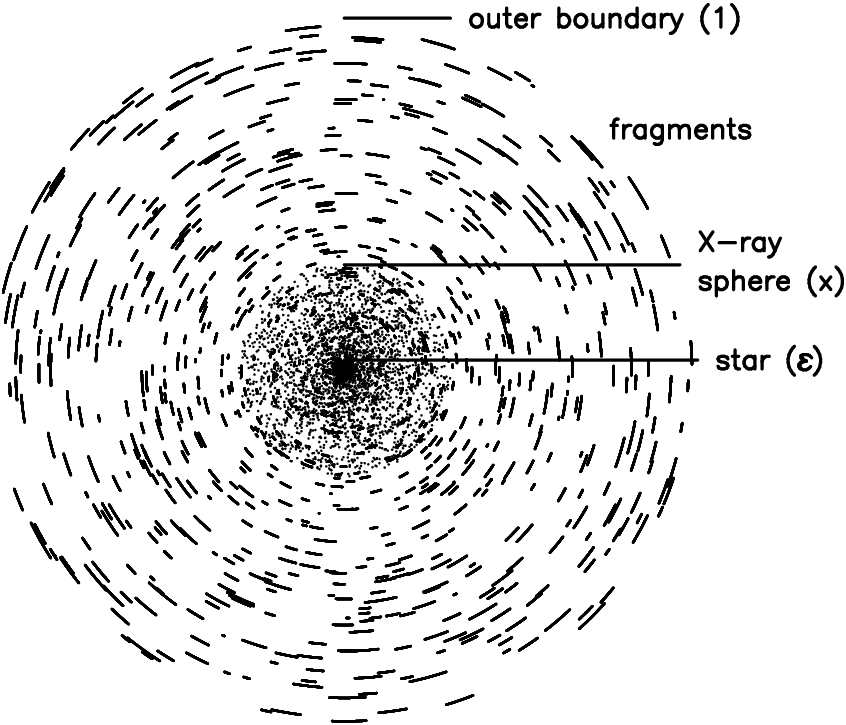}
\caption{Fragmented wind, with $N=10$ fragments per radial ray. The
symbols in brackets designate radii.}
\label{fig_pic}
\end{figure}

Assumptions (1) and (2) are chosen for calculational simplicity. The
present paper focuses on qualitative aspects of X-ray line formation,
and future work shall deal with quantitative modelling of line
profiles. Assumption (3) is the crucial one, of a wind with {\it
oriented} absorber fragments. Consistent with assumption (1), the
fragments partake in the radial wind expansion (assumption 4) and have
uniform radial distribution (assumption 5). Assumption (6) will be
discussed in Sect.\,\ref{sectierer} below. Assumption (7) is chosen to
derive a principal limiting case, and will be dropped in
Sect.\,\ref{nonopaque}. Assumption (8) accounts for the fact that the
X-ray emitting region may be smaller than the region containing
absorbing fragments. Figure\,\ref{fig_pic} shows a sketch of this
fragmented wind model.

\subsection{Photon escape}
\label{sectierer}

We discuss assumption (6) and its relation to the total radial wind
optical depth. The $4\pi$ solid angle as seen from the star is covered
$N$ times by absorbing fragments. One may wonder how photons can
escape from an atmosphere with $N$-fold coverage by opaque absorbers.
This is due to two effects, lateral absorber randomization and
sphericity.

\noindent{\it Lateral randomization.} One easily proofs that covering
an arbitrary planar surface of area $A$ with $N$ thin layers of opaque
absorbing material, then cutting each of the layers into small pieces
and distributing the latter randomly and uniformly within $A$, an area
$Ae^{-N}$ is left empty of absorbers. Therefore, photons crossing
through $A$ experience an optical depth $\tau=N$. For such a laterally
randomized absorber distribution, there are still {\it on average} $N$
absorbers along each ray crossing through $A$, as is specified in
assumption (6) above.

\begin{figure}
\centering
\includegraphics[width=7cm]{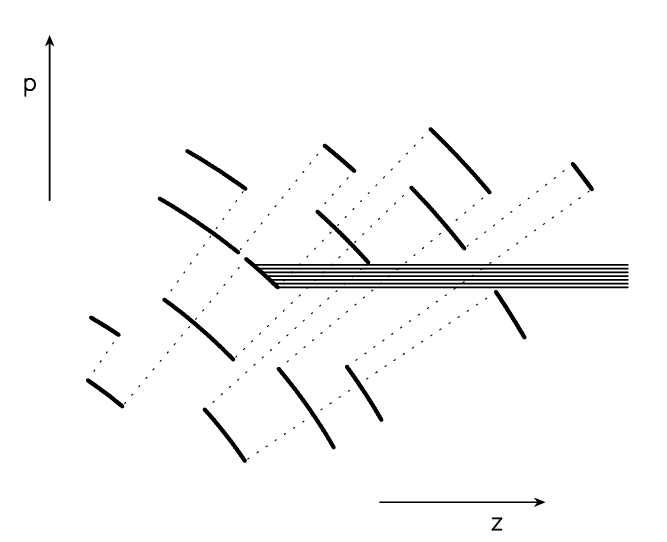}
\caption{Photon escape chanels for $N=3$. The dotted lines indicate
where dense shells got fragmented. The lines drawn parallel to the $z$
axis indicate photon escape chanels towards the observer.}
\label{fig_exa}
\end{figure}

\noindent{\it Sphericity.} We sharpen for the moment assumption (6) to
the condition that there are {\it exactly} $N$ fragments along each
radial ray. This would be the case if the flow is purely radial,
without any lateral motions, as is indicated in Fig.\,\ref{fig_exa}.
Despite the $N$-fold coverage, photon escape chanels occur like the
ones shown in the figure.

Taking assumption (1) of purely radial flow at face value, photons
could {\it only} escape via sphericity. {\it All} central rays from
the origin $r=\reps, p\rightarrow 0$ encounter $N$ opaque fragments,
and the optical depth is $\infty$, not $N$. In the present paper we do
not interpret (1) in this strict sense, but assume that lateral
fragment randomization has occured. This is partly justified by noting
that the random {\it radial} distribution of fragments specified in
assumption (5) appears as {\it lateral} randomization when viewed
under an angle $\theta$. In summary, assumptions (6) and (7) fix the
total optical depth in radial direction at $N$.

\section{Opaque fragments}

\subsection{Optical depth in fragmented wind}
\label{opit}

To calculate the optical depth in radial direction for the fragmented
wind, consider an arbitrary radius $r$, infinitesimal radial increment
$\dd r$, and an area element $A$ of arbitrary shape that is oriented
perpendicular to the radial direction. $A$ shall be much larger than
the average local fragment area. The radial optical depth increment is
(note, $r=1$ at the atmospheric rim),
\begin{equation}
\label{audi}
\dd\tau_r={\dis \dd\! A_b\over \dis A}={\dis NA\,\dd r\over \dis A
(1-\reps)} \approx N\,\dd r,
\end{equation}
where $\dd\! A_b$ is the area of opaque atomic absorbers in the volume
$A\,\dd r$. From Eq.\,(\ref{audi}), the total radial optical depth
between $\reps$ and 1 is $N$, in agreement with the foregoing section.
We drop from now on $\reps$ in nonsingular expressions, still using
equation signs, i.e.~$\reps \rightarrow 0$ is assumed.

\begin{figure}
\centering
\includegraphics[width=5cm]{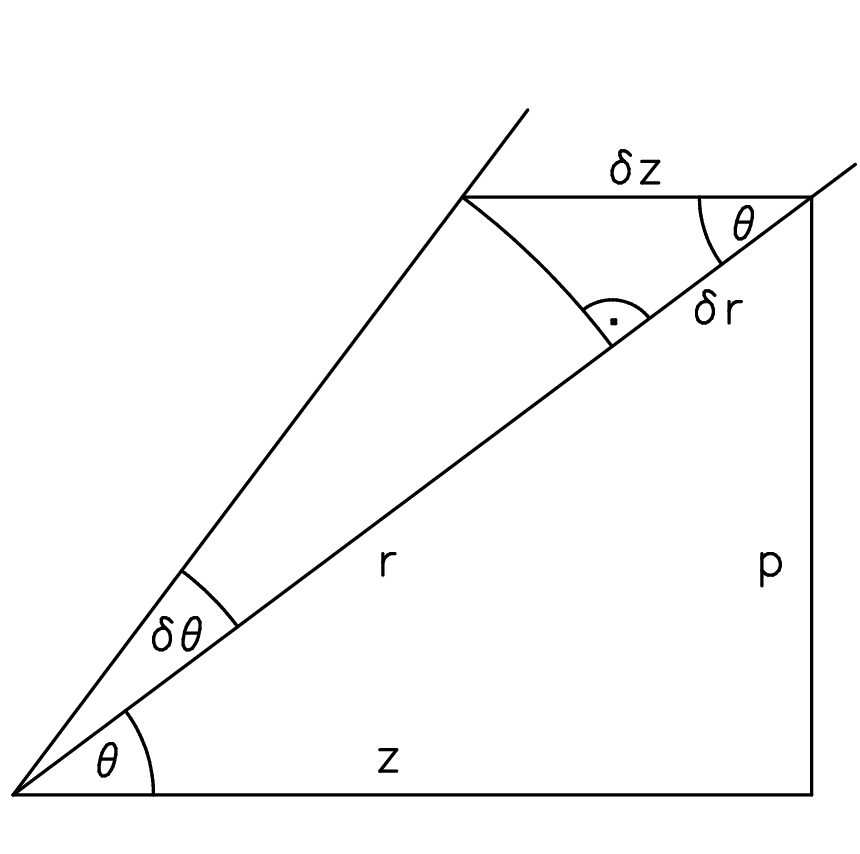}
\caption{Basic geometry in the wind.}
\label{fig_tri}
\end{figure}

The observer shall be located at $z=\infty$ in standard $pz$
coordinates. Introducing the direction cosine $\mu= \vecr\cdot\vecz$
with unity vectors $\vecr$ and $\vecz$, one has $v_z=\mu v_r$ and $\dd
z=\dd r/\mu$, see~Fig.\,\ref{fig_tri}. The optical depth increment for
a photon propagating in $z$ direction is,
\begin{equation}
\label{dtauf}
\dd\tau_z= N \,\dd z\,\mu = N\,\dd r.
\end{equation}
The factor $\mu$ accounts for the reduction in optical depth caused by
the foreshortening of the fragment area in photon propagation
direction. Note that this is equivalent to an angle-dependent opacity
\begin{equation}
\label{wropa}
\chi=\mu N.
\end{equation}
Another way to look at Eq.\,(\ref{dtauf}) is that, for a given pathway
$\dd z$, $N\,\dd r$ is the average number of fragment hits (assuming a
uniform fragment distribution), which is precisely the definition of
the optical depth increment. This is elaborated upon in the
appendix. The total optical depth, $\tau(r,\mu)$, from an X-ray
emission site specified by coordinates $r$ and $\mu$ to the outer
boundary of the atmosphere along the $p$ ray with $p=r\brut$ is
therefore simply the difference of the respective {\it radii $r$} at
these two locations,
\begin{equation}
\label{tauf}
\tau(r,\mu)=
\begin{cases}
N(1-r)    &\text{if $\mu>0$},\\
N(1+r-2p) &\text{if $\mu<0$}.
\end{cases}
\end{equation}
We neglect obscuration due to the central region of radius $\reps$,
which is easy to include in a numerical code but would cause lengthy
case distinctions in the present equations.

The conclusion is that $\tau(r,\mu)$ is independent of $p=p(r,\mu)$
and therefore of $\mu$ if $\mu>0$, i.e.~for the blue part of the line
profile forming in gas that moves towards the observer. Therefore, the
optical depth is the same for all photons that originate on the
surface of a given front hemisphere. The reason is that $\tau$ is the
expectation value of the number of fragment hits between the emission
site and the outer boundary of the atmosphere. Since the spatial
distribution of fragments depends neither on latitude nor azimuth, one
can arrange them, on average, to spherical shells. The same number of
shells, $N(1-r)$, is crossed by each $p$ ray that starts on the
hemisphere with radius $r$. For $\mu<0$, on the other hand, i.e.~the
red part of the profile, the number of shells crossed is $N(r-p)$ on
the back hemisphere, and $N(1-p)$ on the front hemisphere, giving in
total $\tau(r,\mu<0)= N(1+r-2p)$ as in Eq.\,(\ref{tauf}).

\subsection{Optical depth in homogeneous wind}

We compare the above expressions for $\tau$ with those in homogeneous
wind. The fragments consist of atomic absorbers of cross section
$\sigma_0$ that have number density $n_0/r^2$ in the homogeneous wind
in which the fragments form ($v=v_\infty$; and we chose a reference
radius $r_0=1$). The total optical depth of the homogeneous wind in
radial direction is $t= \int_{\reps}^1 \sigma_0 n(r) dr$. In the limit
$\reps \rightarrow 0$, $t=\sigma_0 n_0/\reps$, using equation signs if
errors of order $\reps$ are made. The optical depth increment in $z$
direction is $\dd\tau_z= \sigma_0 n(r) \,\dd z= t\reps \dd z/r^2$.
Note that for isotropic absorbers, no $\mu$ factor occurs here in
contrast to Eq.\,(\ref{dtauf}). According to Fig.\,\ref{fig_tri},
\begin{equation}
\label{consist}
\dd\tau_z=t\reps {\dis\dd\theta\over\dis p},
\end{equation}
where $\dd\theta=\dd z\brut/r$ is the change in latitude over the
pathway $\dd z$. With $p$ being constant, the $z$ integral is replaced
by a $\theta$ integral, giving (e.g.~MacFarlane et al.~1991)
\begin{equation}
\label{tauh}
\tau(r,\mu)= {\dis t\reps\over\dis p} (\acos\,\mu-\asin\, p),
\end{equation}
independent of the sign of $\mu$.

\subsection{Flat-topped profile from emitting sphere in an 
optically thin wind}

Consider a spherical shell with outflow speed $v$ and uniform surface
distribution of X-ray emitters. There shall be no absorption in the
volume inside or outside the shell, and stellar occultation is
neglected. The X-ray intensity measured by an observer at infinity is
$I_\nu=dE_\nu/ d\nu\,d\omega\,dt\,dA$, with the usual meaning of the
symbols. Integrating over {\bf n}~$d\omega$ gives for the flux,
$F_\nu\sim dE_\nu/d\nu$. Due to the Doppler effect, $\nu=\nu_0+\mu
v/v_{\rm th}$, where $\nu_0$ is the rest frame frequency of the line
and $v_{\rm th}$ the thermal speed. Hence $d\nu\sim d\mu$ and
$d\nu/d\theta\sim\sin\theta$. The energy emitted at a given line
frequency $\nu(\mu)$ is $dE_\nu\sim \eta\,dV= \eta\,dr\cdot
rd\theta\cdot 2\pi r\sin\theta$. Therefore $F_\nu\sim dE_\nu/d\theta
\cdot d\theta/d\nu \sim r^2 dr\,\eta(r)$. Since any reference to $\mu$
(or $\theta$) has vanished, the line profile is flat-topped.

\subsection{X-ray line profiles for opaque fragments}

We turn to X-ray line formation in a dense wind with absorption. For
constant radial wind speed $v(r)=v_\infty$, the surfaces of constant
projected velocity $\mu v$ as seen by an observer at infinity are
straight cones of opening angle $\acos\,\mu$. The emergent line
profile is given by
\begin{equation}
F^e_\mu\sim\int_{\reps}^{\rx} dr\,r^2\,\eta(r)\,
e^{-\tau(r,\mu)},
\end{equation}
with $\tau(r,\mu)$ from Eqs.\,(\ref{tauf}) or (\ref{tauh}). Only
$\tau$ depends on $\mu$, i.e.~refers to the explicit cone considered.
Since $\tau(r,\mu)$ from Eq.\,(\ref{tauf}) is {\it independent} of
$\mu$ for $\mu>0$, the blue part of the line profile from the
fragmented wind remains flat-topped, despite absorption. This is {\it
not} true for the homogeneous wind. All photons emitted from the
surface of a given front hemisphere are diminished by the same optical
depth (the fragment hit expectation value), hence the emission line
profile stays flat-topped even after severe absorption. Note that in a
wind with nonconstant outflow speed, the flat-topped profiles from
individual emitting spheres of radius $r$ have different widths
$[-v(r),+v(r)]$ in velocity space, hence the sum of individual line
profiles is no longer a flat-topped profile.

We repeat the physical argument for flat-topped blueshifted parts of
line profiles despite wind absorption: divide the X-ray emitting
volume into spherical layers of infinitesimal thickness. If the wind
would be optically thin, the line emission from each of these shells
would show a flat-topped profile. Consider the front hemisphere
towards the observer where the blueshifted component of the profile
forms. Because the distribution of absorbing fragments obeys spherical
symmetry, one can arrange them, on average, on spherical shells (with
uniformly distributed holes). Therefore, all photons reaching the
observer at infinity from a given radius $r$ in the front hemisphere
encounter, on average, the same number of absorber fragments,
i.e.~experience the same optical depth. Hence, the blue part of the
profile remains flat.

\begin{figure}
\centering
\includegraphics[width=6cm]{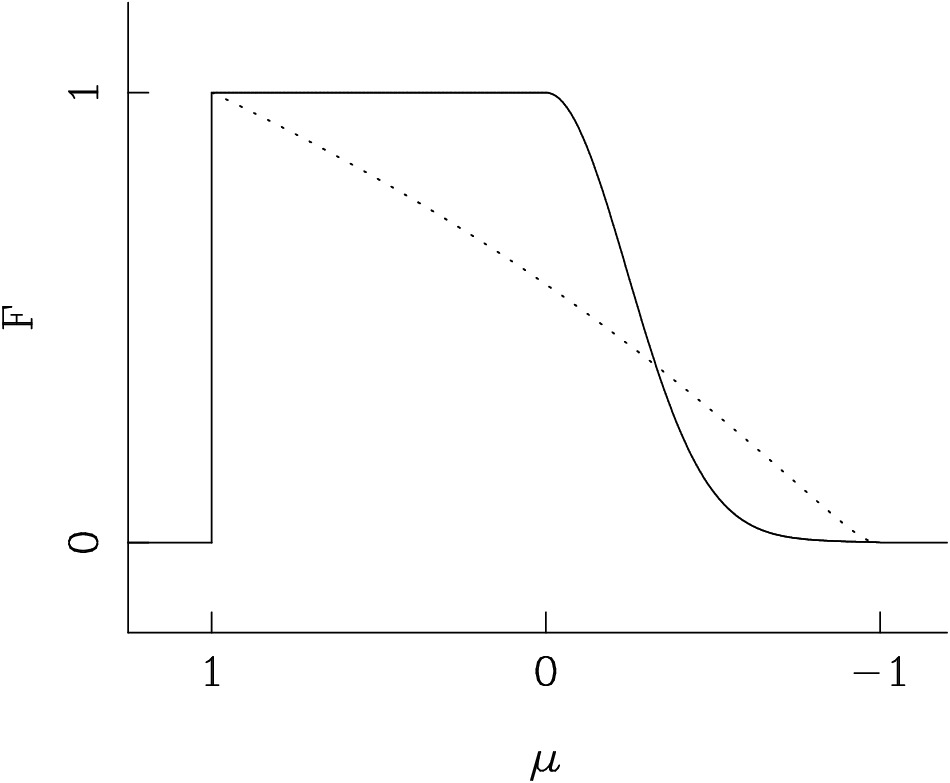}
\caption{X-ray emission line profiles. Full: fragmented wind with
opaque absorbers; dotted: homogeneous wind. Parameters are: $N=10$
dense fragments per radial ray; inner atmospheric radius $\reps=0.1$;
termination radius $\rx=1$ for X-ray emission; radial optical depth
$t=3$ of the homogeneous wind. The abscissa is in units of the
projected, normalized wind speed, $\mu=v_z/v_\infty$.}
\label{fig_pr1}
\end{figure}

Figure\,\ref{fig_pr1} shows line profiles from the fragmented and, for
comparison, from the homogeneous wind. The latter shows the well-known
red-blue asymmetry, with a rather uniform decline from the blue to the
red line wing (MacFarlane et al.~1991; Ignace 2001; Owocki \& Cohen
2001). By contrast, the profile for the fragmented wind is almost a
step function, namely flat-topped over the whole blueshifted part, and
with steep decline to zero intensity on the redshifted part. We
suggest that such a profile, after convolution with the detector
response function, resembles the blueshifted yet symmetric emission
line profiles observed from $\zeta$~Pup, see Fig.\,3 in Cassinelli et
al.~(2001). By the same token, its blueshift is too large to be
reconcilable with $\zeta$~Ori's observed emission lines. For Chandra
and XMM-Newton, the line center frequency can be measured with an
accuracy of roughly 5\,m\AA, which corresponds to $\dd v/v_\infty
\approx 0.06$ at 1 keV, assuming $c/v_\infty=150$. This is almost one
order of magnitude smaller than the blueshift $\dd v/v_\infty\approx
0.5$ in Fig.\,\ref{fig_pr1}.

As a conclusion from this section, observed X-ray line profiles of
$\zeta$~Pup and $\zeta$~Ori are qualitatively different. Our model of
a fragmented stellar wind can only explain the former.

\section{Nonopaque fragments}
\label{nonopaque}

\subsection{Fragment optical depth}

We drop now assumption (7), and account for the finite optical depth
of fragments. The latter shall form close to the photosphere in a
rather uniform process, and therefore contain roughly the same number
of atoms. The radial optical depth of a fragment is the absorber area
per fragment area, $\sigma_0 n_0/Nr^2$ or $t\reps/Nr^2$. In the limit
$1/N \equiv dr \rightarrow 0$ of infinitely many, optically thin
fragments, the integral over $dr$ gives $t$, as must be. On the other
hand, for a finite number $N$ of optically thick fragments, the
expression for the total radial optical depth becomes,
\begin{equation}
\label{tinte}
\tf=\int_\reps^1 dr\,N\, \left[1-\exp\left({-t\reps\over
Nr^2}\right)\right].
\end{equation}
This is derived in Eq.\,(\ref{ende2}) of the appendix, but is also
intuitively clear: the integrand is the product of the hit expectation
value of a fragment -- which is the optical depth increment for opaque
fragments, see Eq.\,(\ref{audi}) -- times the transmission probability
for nonopaque fragments. For a large number of optically thin
fragments, $N\rightarrow\infty$, the exponential can be Taylor-series
expanded to first order, giving again $\tf=t$. In the opposite limit
of opaque fragments, the exponential term is dropped, giving $\tf=N$
as in Sect.\,\ref{sectierer}. Turning to moderately optically thick
fragments, we shift the lower integration bound in Eq.\,(\ref{tinte})
to 0 and make a variable substitution $y=1/r$. Integrating by parts,
\begin{equation}
\tf/N=1 - e^{-a} + \sqrt{a\pi}\bigl(1-{\rm erf}\sqrt{a}\bigr),
\end{equation}
where $a=t\reps/N$ and ${\rm erf}\,x=2\pi^{-1/2}\int_0^x e^{-t^2}
dt$. The average optical depth per fragment is $\tf/N=0.17$ for
$a=0.01$ (optically thin fragments), is 0.46 for $a=0.1$, and is 0.91
for $a=1$ ($\tf/N=1$ for opaque fragments).

Cassinelli \& Olson (1979) and Hillier et al.~(1993) estimate that the
total radial optical depth of the wind of $\zeta$~Pup could reach 30
at 1~keV. Time-dependent hydrodynamical simulations of O~star winds
suggest that there are of order 10 dense shells. Assuming, then,
$t=30$, $N=10$, and $\reps=0.1$, one has $t\reps/N=0.3$. Around this
value of $t\reps/N$, the line profile shape starts to resemble the
profile from opaque fragments, see Figs.\,\ref{fig_pr2} and
\ref{fig_pr3}. We must thus conclude that assumption (7) of opaque
fragments is only applicable for rather dense winds.

\subsection{X-ray lines profiles for nonopaque fragments}

According to the above, the optical depth $\tau_{r\mu}$ of a single
fragment located at radius $r$, with direction cosine $\mu$, is
\begin{equation}
\label{fino}
\tau_{r\mu}={t\reps\over Nr^2\mu}
\end{equation}
in $z$ direction. Here, $\mu$ accounts for the longer pathway through
the fragment when seen under an angle $\theta$. The formal integral
for the line profile shape becomes
\begin{align}
\label{profile2}
F^e_\mu &\sim \int_{\reps}^{\rx} dr\,r^2\,\eta(r)\, e^{-\tau(r,\mu)},
\notag\\
\tau(r,\mu)&=\int_r^1 dr' \,N\, \left[1-\exp\left({-t\reps\over
Nr'^2\mu'} \right)\right].
\end{align}
This cannot be solved analytically, but is easily evaluated
numerically.

\begin{figure}
\centering
\includegraphics[width=6cm]{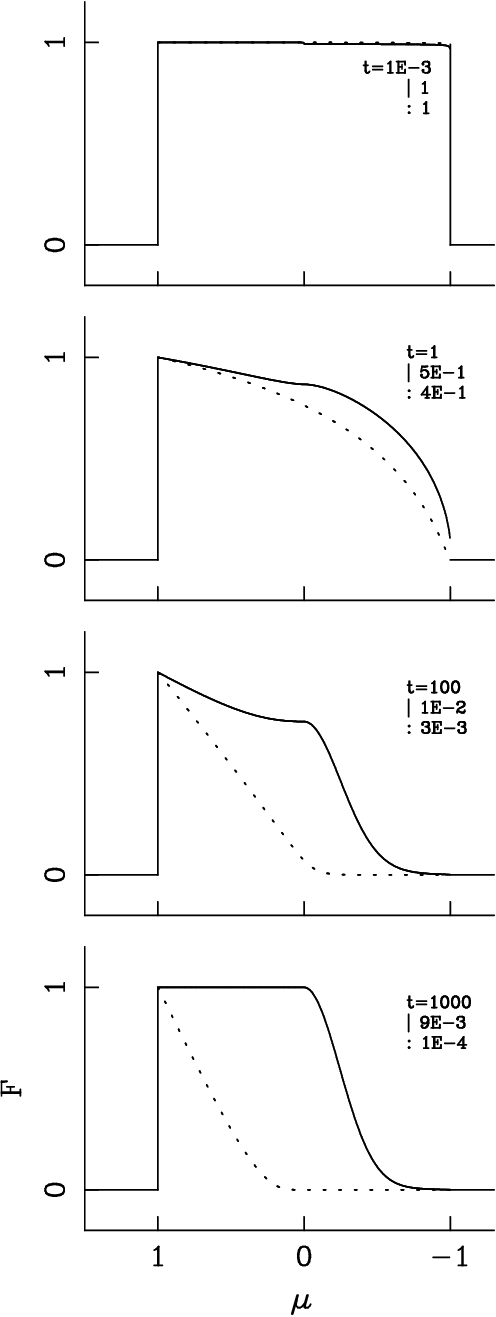}
\caption{Normalized X-ray emission line profiles from winds with
increasing optical depth (top: optically thin; bottom: very optically
thick). Full lines show the fragmented wind model, dotted lines the
homogeneous wind. Parameters are $N=10$, $\reps=0.1$, $\rx=1$,
$\eta\sim r^{-4}$. The numbers at the top right give the total radial
optical depth $t$ of the homogeneous wind, and the ratios of emergent
to intrinsic X-ray flux.}
\label{fig_pr2}
\end{figure}

Figure\,\ref{fig_pr2} shows X-ray line profiles at different values of
the wind optical depth $t$. From Eq.\,(\ref{tinte}), assumption (7) of
opaque fragments holds if $t\reps/N\gap 1$. In accordance with this,
the line profile in the bottom panel of Fig.\,\ref{fig_pr2}, at
$t\reps/N=10$, resembles that for fully opaque fragments from
Fig.\,\ref{fig_pr1}. The same is essentially true for the second panel
from bottom, at $t\reps/N=1$. At smaller $t$, however, the profiles
for the fragmented and homogeneous wind become similar, as is seen in
the two top panels of the figure.

At present it is not clear, which fraction of intrinsic X-rays become
visible as emergent X-rays. For O~stars, the latter typically amount
to $10^{-7} \,L$, with bolometric luminosity $L$ (Sciortino et
al.~1990). The wind kinetic energy, on the other hand, which is the
ultimate reservoir for X-ray energy via thermalization of macroscopic
flow energy into heat by shocks, comprises for $Lv_\infty/2c$ or
$10^{-3}-10^{-2}\,L$ if the wind is close to the single-scattering
limit, $L/c\approx \dot M v_\infty$, and if $v_\infty \lea c/100$.
Therefore, there is a large potential margin between intrinsic and
emergent X-rays. The models shown in Fig.\,\ref{fig_pr2} all lie
within this margin.

\subsection{Size of X-ray emitting region}
\label{xsize}

\begin{figure}
\centering
\includegraphics[width=6cm]{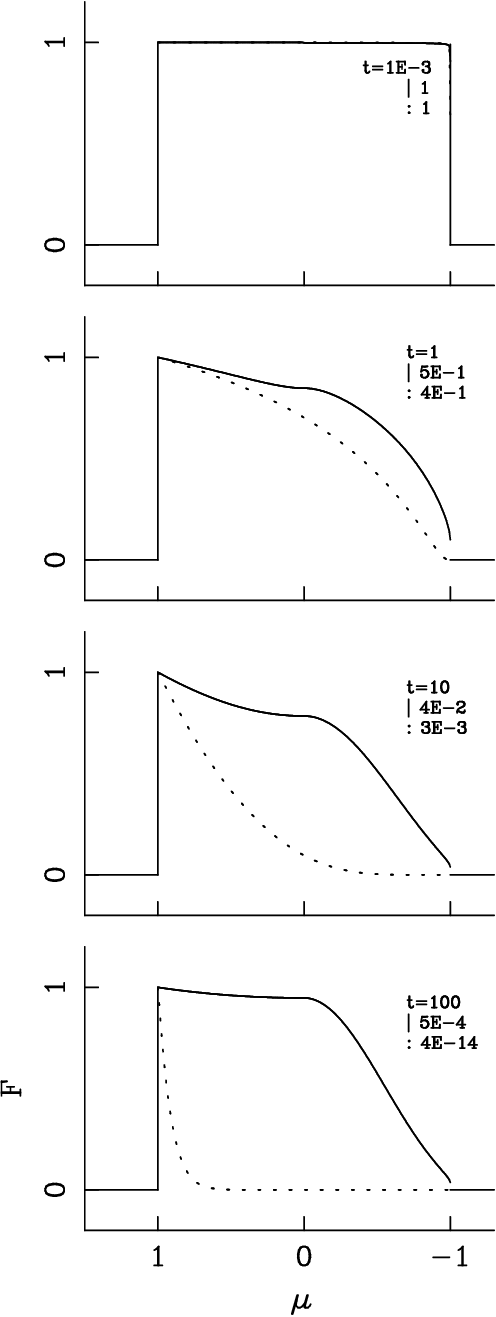}
\caption{Similar to Fig.\,\ref{fig_pr2}, but with smaller X-ray
emitting volume, $\rx=0.3$. Still $N=10$, $\reps=0.1$.}
\label{fig_pr3}
\end{figure}

Hydrodynamic simulations show that shock-compressed, absorbing
fragments significantly reexpand above $\approx 30\,R_\ast$ due to
internal pressure, and the wind gradually becomes homogeneous
again. On the other hand, X-ray emitting shocks (e.g.~as caused by
collisions of fast wind clouds with fragments) may only reach out to
$10\,R_\ast$, suggesting $\rx=0.3$ in assumption (8).

Figure\,\ref{fig_pr3} shows line profiles for this value of $\rx$. The
major differences when compared to Fig.\,\ref{fig_pr2} are: (i) the
decline on the red part of the profile is more gradual now; (ii) the
blue part of line profiles from the fragmented wind starts to become
flat-topped at $t\reps/N \gap 0.1$ already; (iii) for the homogeneous
wind at $t\reps/N=1$, the ratio of emergent to intrinsic X-rays is
very small, $4\times 10^{-14}$.

\subsection{Natal fragment absorption}
\label{natal}

We assumed so far that X-rays originate from {\it homogeneous}
emission within a sphere of radius $\rx$. Instead, X-ray emission may
spatially coincide with the oriented, absorbing fragments. The line
deshadowing instability is expected to form strong, X-ray emitting
shocks that cool radiatively, leading to highly compressed fragments
of cool, absorbing wind gas. If the radiative cooling zone is narrow,
one can think of both X-ray emission and absorption occuring in the
same, spherical segments. The shocks are of the braking type, i.e.~are
reverse shocks on the starward facing side of fragments. X-ray
emitters on the front hemisphere, $\mu>0$, are then hidden behind
their natal fragment, whereas emitters on the back hemisphere are
not. To treat natal fragment absorption, we have to account for the
finite optical depth of fragments, since else the whole blueshifted
part of the line profile would turn black. For photons from the back
hemisphere, we use $\tau(r,\mu)$ from Eq.\,(\ref{profile2}) in the
formal integral, however with $N\,dr$ replaced by $(N-1)\,dr$ because
the natal shell is not hit. For photons emitted from the front
hemisphere, $\tau(r,\mu)$ is replaced by $\tau(r,\mu)+\tau_{r\mu}$,
with $\tau_{r\mu}$ the optical depth of the natal fragment from
Eq.\,(\ref{fino}). In Eq.\,(\ref{profile2}), $N\,dr \rightarrow
(N-1)\,dr$ again, to avoid double counting the optical depth of the
natal shell.

\begin{figure}
\centering
\includegraphics[width=6cm]{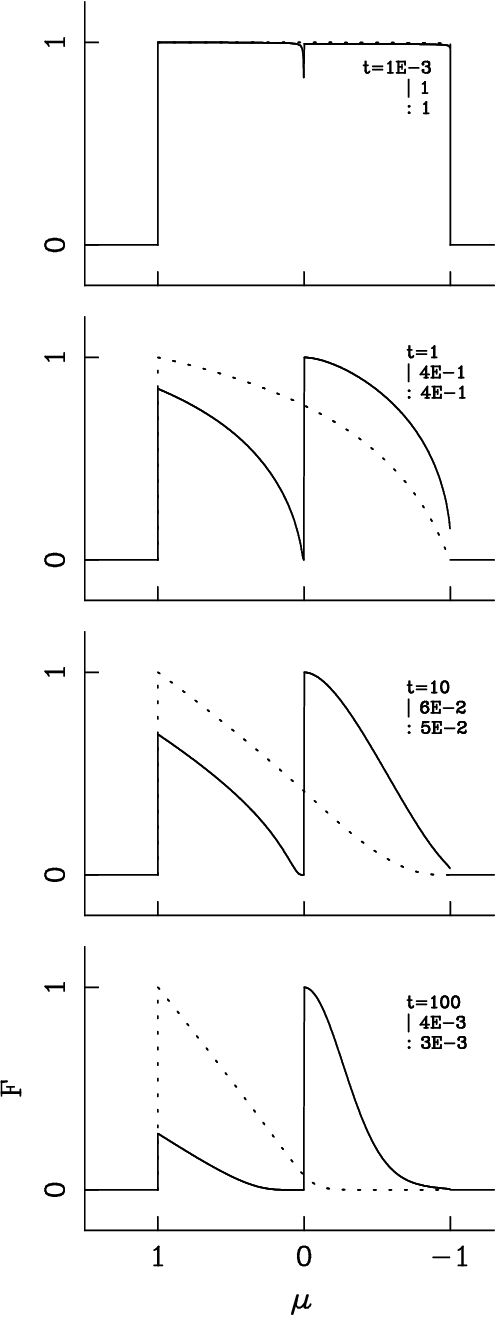}
\caption{Similar to Fig.\,\ref{fig_pr2}, but with natal fragment
absorption included.}
\label{fig_pr4}
\end{figure}

Figure\,\ref{fig_pr4} shows the resulting line profiles. Since
$\tau_{r\mu}$ from Eq.\,(\ref{fino}) becomes infinite at $\mu=0$, the
center of the profile from the optically thick, fragmented wind is
black. This kind of double-peaked X-ray emission line profile is not
observed. The resolution of the Chandra and XMM-Newton gratings is
sufficiently high as to resolve such a profile shape.

We are drawn to conclude that either forward shocks exist of a
strength similar to that of the reverse shocks, and prevent the drop
to zero intensity; or that the assumption of spatial proximity of
X-ray emitters and absorbers in 2-D, oriented fragments is too
strong. The former alternative seems unlikely; the latter is rather
plausible, given that both lateral motions (eddies) and a finite
extent of radiative cooling zones could prevent emitters at $\mu\ge 0$
from being hidden behind their natal fragment. Therefore, a
homogeneous distribution of X-ray emitters as in assumption (8) seems
more appropriate than a strict coincidence of emitters and absorbers
in 2-D, oriented fragments.

\section{Summary}

We consider X-ray emission line formation in a spherically symmetric
stellar wind with constant radial outflow speed. X-rays are assumed to
originate from hot gas within a sphere of radius $\rx$. Cool,
absorbing wind gas resides in flat, highly overdense fragments that
are oriented perpendicular to the radial flow direction. The fragments
are thought to result from radial fragmentation of compressed gas
shells that form due to the line deshadowing instability. The
fragments have infinitesimal opening angle, a uniform spatial
distribution, and partake in radial expansion. Our results are:

1.~For opaque fragments, the blueshifted part of the line profile is
flat-topped, even after strong absorption, whereas the redshifted part
drops steeply to zero. The overall profile is therefore essentially
flat-topped with a full width of $v_\infty$ and a blueshift of
$v_\infty/2$. This is in stark contrast to the well-known gradual
decline from the blue to the red line wing for a homogeneous wind. The
line profiles from the fragmented wind could explain why observed
X-ray line profiles from $\zeta$~Pup are strongly blueshifted, yet can
be fitted by a Gaussian function.

2.~A blueshift of $v_\infty/2$ is too large to be in accord with the
(essentially) unshifted X-ray line profiles observed from $\zeta$~Ori
and $\theta^1$~Ori~C. We suggest that X-rays from these stars do not
originate in instability-generated wind shocks, but in colliding wind
shocks or processes involving magnetic fields.

3.~The assumption of opaque fragments is applicable if $t\reps/N\gap
1$ ($t$ the radial optical depth of the homogeneous wind between radii
$\reps$ and 1; $N$ the number of fragments along radial rays), as
could be the case for dense O~star winds like $\zeta$~Pup's. On the
other hand, with decreasing optical depth of nonopaque fragments, line
profiles become similar to those from a homogeneous wind.

4.~X-ray absorption in natal fragments leads to zero flux at line
center. This kind of double-peaked profile is not observed, which
suggests that lateral motions and a finite extent of radiative cooling
zones prevents a strict spatial coincidence of X-ray emitters and
absorbers in 2-D, oriented fragments.

Quantitative modelling of X-ray emission line profiles of $\zeta$~Pup
accounting for an accelerating wind and instrumental broadening will
be subject of a forthcoming paper.

\begin{acknowledgements}
We thank Joachim Puls and Stan Owocki for helpful discussions, and the
anonymous referee for suggesting to include the case of nonopaque
fragments. This work was supported by the Deut\-sche
For\-schungs\-ge\-mein\-schaft under grant Fe~573/1-1.
\end{acknowledgements}

\appendix
\section{A Poisson ring}

We take here a different view at X-ray absorption in a wind with
oriented absorbers, using Poisson statistics. The two objectives of
this appendix are (i) to make explicit the connection of fragment hits
and optical depth, and (ii) to obtain an expression for X-ray
transmission through nonopaque absorbers. We assume $N=1$ throughout,
and generalize to $N>1$ at the end of the appendix. In the fragmented
wind, consider a ring with radius $p$ centered on the $z$ axis at
latitude $\theta$. The probability that a photon propagating at impact
parameter $p$ hits an absorber fragment along the ring is $\dd w_z=\dd
r$ within a pathway $\dd z$. Decrease $\dd z$ until each fragment that
is hit along the ring within this $\dd z$ {\it fully} covers the
interval $[\theta, \theta+\dd\theta]$. Next, divide the ring into
$N_\phi$ azimuthal segments of width $\dd\phi=\dd\theta$. Since the
fragment hit probability $\dd w_z$ in each of these segments is small
and the number $N_\phi$ of segments (or ``trials'') is large, the
requirements for a Poisson distribution are met. The single parameter
of the latter is the expectation value, $\lambda$, here of {\it
fragment hits},
\begin{equation}
\lambda=N_\phi\,\dd w_z= {\dis 2\pi\mu p\over\dis\bruts},
\end{equation}
using $\dd r=\mu p\,\dd\phi/(\bruts)$. To give an example, consider
the ring with $p=1/2$ and $\theta=45^\circ$ ($\mu=1/\sqrt{2}$).  On
average, $\lambda= \pi\sqrt{2}\approx 4.44$ fragments are hit along
this ring, in a pathway $\dd z$. The independence of the number of
hits from $N_\phi$ is easily understood: if $N_\phi$ is doubled,
$\dd\phi$, $\dd\theta$, $\dd z$, and $\dd w_z$ drop by a factor of
two, leaving $\lambda$ unchanged.

The probability $\pi(k)$ that $k$ fragments are hit along the full
ring is,
\begin{equation}
\label{poisson}
\pi(k)={\dis\lambda^k\over \dis k!} e^{-\lambda},\qquad k=0,1,2,\ldots
\end{equation}
Photons at different $\phi$ have different absorption histories, but
for simplicity we assume that the whole ring is illuminated by an
average intensity $\Ii$. Summing the intensity $I^{\rm o}$ after
absorption within $\dd z$ over all segments $\dd\phi$,
\begin{align}
&I_1^{\rm o}+I_2^{\rm o}+\ldots I_{N_\phi}^{\rm o} = N_\phi \Io
=\hfill\notag\\
&\qquad = \pi(0) \;\Ii\; N_\phi \notag\\
&\qquad + \pi(1) \;\Ii\; \bigl( N_\phi-1 +  e^{-\tau_{r\mu}}\bigr) \notag\\
&\qquad + \pi(2) \;\Ii\; \bigl( N_\phi-2 + 2e^{-\tau_{r\mu}}\bigr) \\
&\qquad + \ldots \notag\\
&\qquad + \pi(N_\phi) \;\Ii\;\; N_\phi \,e^{-\tau_{r\mu}},\notag
\end{align}
where $\tau_{r\mu}$ is the fragment optical depth at radius $r$ in
direction $\acos\,\mu$. Inserting (\ref{poisson}) and using the
binomial formula for $k\ll N_\phi$ (as can be safely assumed for a
Poisson distribution), it follows that
\begin{equation}
\Io=\Ii \, e^{-\lambda} \sum_{k=0}^\infty {\dis\lambda^k \over \dis k!}
   \left(\dis 1-{1-e^{-\tau_{r\mu}}\over \dis N_\phi}\right)^k.
\end{equation}
The sum is the power series expansion of the exponential,
\begin{equation}
\label{ende}
\Io=\Ii\; e^{-\lambda} e^{\lambda \left(1- {1-e^{-\tau_{r\mu}}\over
   N_\phi}\right)} =\Ii\; e^{-\dd w_z\left(1-e^{-\tau_{r\mu}}\right)}.
\end{equation}
The case $N>1$ is treated by making $N$ copies of the atmosphere, and
eliminating $N-1$ fragments along the radial rays in each of the
copies. The $N$ copies with one remaining fragment per radial ray are
arranged like pearls on a string, and photons pass subsequently
through each of them. Hence, for $N>1$, $\delta w_z$ in
Eq.\,(\ref{ende}) has to be replaced by $N\,\dd w_z$, giving as
optical depth increment along $\dd z$,
\begin{equation}
\label{ende2}
\dd\tau_z=N \left(1-e^{-\tau_{r\mu}}\right)\dd r.
\end{equation}
For opaque fragments, this reduces to Eq.\,(\ref{dtauf}) of the main
text. According to Eq.\,(\ref{ende2}), the partial transparency of the
fragments leads to a reduction of their effective opacity by a factor
$1-\exp(-\tau_{r\mu})$, i.e.~we obtain
\begin{equation}
\chi=\mu N \left(1-e^{-\tau_{r\mu}}\right)
\end{equation}
in replacement of Eq.\,(\ref{wropa}).

\end{document}